# Wave Propagation in Porous Elastoplastic Rocks: Implication for Seismic Attenuation

*Viktoriya Yarushina*[1], *Alexander Minakov*[2]

[1]Institute for Energy Technology, Oslo, Norway

[2]University of Oslo, Norway

**Introduction**

The effect of material non-linearity on seismic wave propagation is well-documented by observations and experiments [Guyer and Johnson, 2008]. Non-linearity of seismic wave propagation is important for understanding various natural phenomena in dynamic porous media including wave energy attenuation, wave-induced permeability enhancement, and soil liquefaction. We suggest that this nonlinearity emerges from the nonlinearity of the geomechanical behavior inherent to the fluid-bearing rocks.

According to the mechanism of P-wave attenuation suggested by Yarushina and Podladchikov [2010], seismic energy losses can be attributed to irreversible plastic yielding. It has been shown that small amplitude seismic waves, propagating through the pre-stressed rocks, can cause nonelastic strain resulting in attenuation of some fraction of the wave energy per loading cycle. The microscale plastic yielding gives rise to frequency-independent attenuation due to rate-independence of the plasticity formulation. We implement this mechanism in a 1D poroacoustic wave propagation code including dynamic porosity evolution and nonlinear rheology.

We test an idea that the interaction of sound waves with a local heterogeneity, residing at abnormal pressure, has a potential to attenuate seismic waveforms. The model is applied to analyze the seismic signatures of gas chimneys. Recently, gas chimneys have attracted a lot of attention from both industry and academia due to their potential as drilling hazards and preferential leakage pathways in hydrocarbon reservoirs and $CO_2$ storage sites. Geomechanical modelling suggests that chimney formation may be associated with complex physical properties that are not described by simple elastic models [*Raess et al.*, 2014]. Thus, accurate computation of synthetic seismograms for media with complex rheology is required for extraction of valuable information. The forward modeling method that correctly describes physics of the wave propagation in such media is the key ingredient of the seismic inversion and imaging techniques.

**Model formulation**

We begin with the analysis of P-wave. Our physical model of seismic wave propagation includes complex two-phase physics and development of irreversible elastoplastic deformations in the rock matrix associated with chimney formation [*Yarushina and Podladchikov*, 2015]. It represents a generalization of Biot's poroelasticity and includes the force balance equations in the form (see Tables 1 and 2 for notations used):

$$\rho^s (1-\varphi) \frac{d^s v_i^s}{dt} + \rho^f \varphi \frac{d^f v_i^f}{dt} = -\nabla_i \bar{p} - g_i \bar{\rho} \qquad (1)$$

$$\rho^f \varphi \frac{d^f v_i^f}{dt} = -\frac{\varphi \eta_f}{k} q_i^D - \varphi \nabla_i p^f - g_i \rho^f \varphi \qquad (2)$$

Mass balance equations in the form:

$$\frac{1}{\rho^s}\frac{d^s \rho^s}{dt} - \frac{1}{1-\varphi}\frac{d^s \varphi}{dt} + \nabla_j v_j^s = 0 \qquad (3)$$

$$\frac{\varphi}{\rho^f}\frac{d^f \rho^f}{dt} + \frac{d^s \varphi}{dt} + \varphi \nabla_j v_j^s + \nabla_j q_j^D = 0 \qquad (4)$$

And compaction equations in the form:

$$\frac{d^s \overline{p}}{dt} = -K_u \left( \nabla_k v_k^s + B \nabla_k q_k^D \right) \qquad (5)$$

$$\frac{d^f p_f}{dt} = -B K_u \left( \nabla_k v_k^s + \frac{1}{\alpha} \nabla_k q_k^D \right) \qquad (6)$$

$$K_\varphi \frac{d\varphi}{dt} = \frac{dp^f}{dt} - \frac{d\overline{p}}{dt} \qquad (7)$$

Rate equations (5) and (6) represent generalization of linear Biot's equations with $\nabla_k v_k^s$ being the rate of volumetric deformation and $\nabla_k q_k^D$ being a rate analogue of the fluid mass content. Equation (7) gives porosity dynamics due to compaction, which was ignored in original Biot's equations. As inferred from the micromechanical modelling for elastoplastic porous rocks [*Yarushina and Podladchikov*, 2010], effective bulk modulus is the following function of the effective stress:

$$K_\varphi = \begin{cases} \dfrac{G}{\varphi} & \text{if} \quad |p_e| < Y \text{ or } \text{sign}(p_e) dp_e \leq 0 & (\textit{elastic}) \\ \dfrac{G}{\varphi} & \text{if} \quad \text{sign}(p_e) dp_e \leq 0 & (\textit{unloading}) \\ \dfrac{G}{\varphi} \exp\left(1 - \dfrac{|p_e|}{Y}\right) & \text{if} \quad |p_e| \geq Y \text{ and } \text{sign}(p_e) dp_e > 0 & (\textit{plastic}) \end{cases} \qquad (8)$$

where yield strength $Y = \max\{|p_e|, Y_0\}$ accounts for material hardening ($Y_0$ is an initial yield strength). Other compaction parameters introduced in equations (5) and (6) depend on the fluid bulk modulus $K_f$, solid bulk modulus $K_s$ and effective bulk modulus $K_\varphi$. Undrained bulk modulus is $K_u = \dfrac{K_d}{1-\alpha B}$; $\alpha$ is the Biot-Willis parameter; $B$ is the Skempton's coefficient; $K_d$ is the drained bulk modulus.

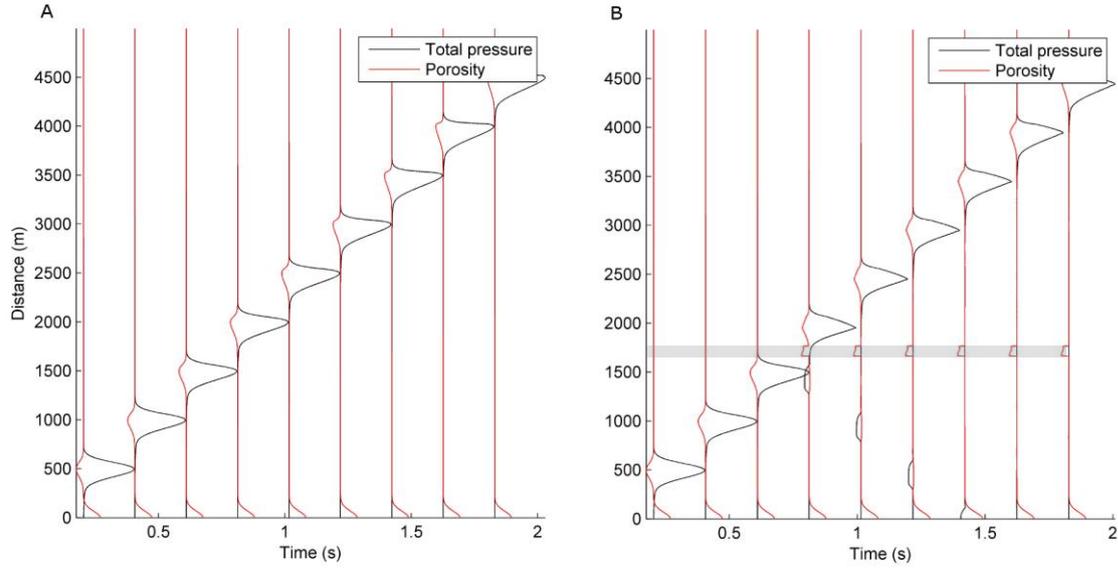

*Figure 1* (A) Acoustic waveforms in a drained porous elastic (A) and porous elastoplastic (B) rocks. Total pressure and porosity are presented. A plane acoustic wave hits a 100-m thick weak layer at the distance of 1667-1767 m.

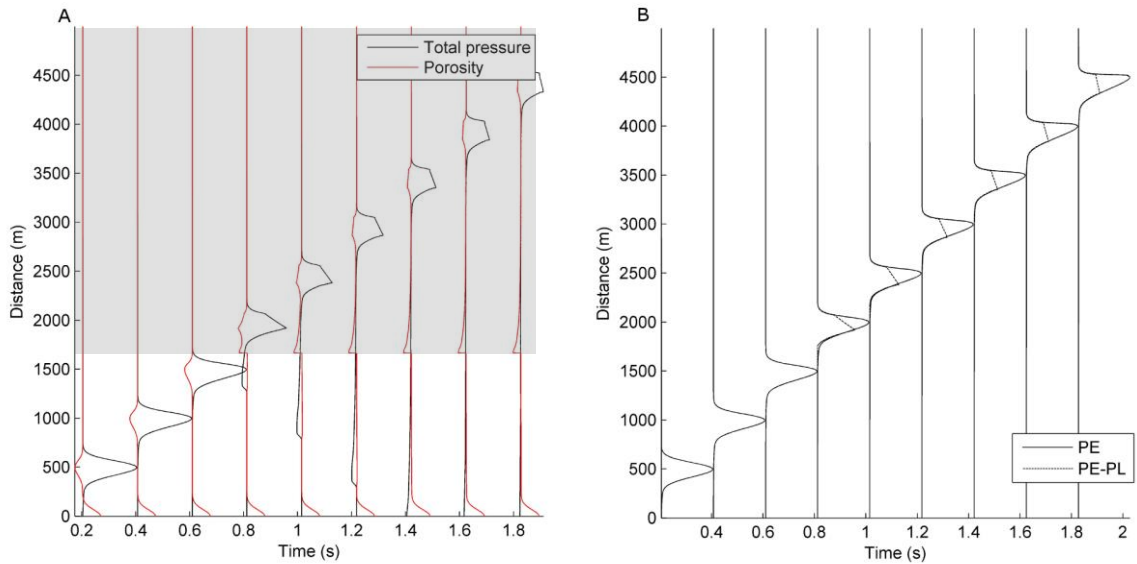

*Figure 2* (A) Acoustic waveforms (total pressure and porosity) in a drained porous elastoplastic rock. A plane acoustic wave hits a weak half space at the distance of 1667 m. (B) Comparison of poroelastic (PE) and poroelastic-plastic (PE-PL) transmitted waveforms (total pressure). The comparison shows the effect of energy attenuation due to plastic yielding.

**Numerical implementation of model equations**

The equations (1) – (8) are solved using 1D finite difference scheme on a staggered grid where pressure and velocity variables are attributed to different grid nodes. An explicit Euler scheme with adaptive time step is used for time integration. Nonlinear iterations of equations (5) - (8) are implemented to obtain correct pressures and porosity at each time step. We present in Fig. 1 a model setup with a 100-m thick layer, characterized by reduced yield stress or lower effective pressure. A similar model is shown in Fig. 2 but the layer is replaced by a weak half space. Our preliminary numerical results show that even for the simplest drained poroelastic conditions, compaction of porous rock introduces nonlinearity into acoustic equations that leads to the sharpening of the wave front and possible development of an acoustic shock if the initial wave amplitude is above a certain critical threshold depending on the properties of the rock (Fig. 1A). Addition of elastoplastic deformation further changes the waveform so that strong asymmetry and cusped crest points are developed (Fig. 1B). Following the initial compaction wave a decompaction wave is induced (Fig. 2A). The wave amplitude attenuation due to plastic yielding is shown in Fig. 1B and Fig. 2B for the yield strength to the wave amplitude ration of 1/3.

**Conclusions**

In this paper, we develop and study a 1D model for the acoustic wave propagation with two-phase physics and irreversible elastoplastic deformations in the rock matrix. We address the effect of the P-wave energy attenuation due to pore-scale plastic yielding in pre-stressed sedimentary rocks. The numerical examples are presented for drained rocks that capture major physical aspects of the process. However, the effects associated with viscous pore fluids are also important and will be presented elsewhere. We anticipate that our model can be used for monitoring of fluid flow in natural and artificial reservoirs using seismic data as well as it can be useful for earthquake engineering.

**Acknowledgements**

AM acknowledges Det Norske Videnskaps Akademi, Statoil, and VISTA (project number 6264) for financial support.

Table 1. List of principal notation

| Symbol | Meaning | Unit |
|---|---|---|
| $g_i$ | Gravitational force | m s$^{-2}$ |
| $G$ | Elastic shear modulus of solid mineral grains | Pa |
| $K_s, K_f$ | Solid and fluid bulk moduli | Pa |
| $K_d, K_u$ | Drained and undrained bulk moduli | Pa |
| $K_\varphi$ | Effective bulk modulus of pore space | Pa |
| $p^s, p^f$ | Pressure of solid and fluid phases | Pa |
| $v_i^s, v_i^f$ | Solid and fluid velocity | m s$^{-1}$ |
| Y | Yield stress/cohesion | Pa |
| $\delta_{ij}$ | Kronecker-delta | |
| $\varphi$ | Porosity | |
| $\rho^s, \rho^f$ | Solid and fluid density | kg m$^{-3}$ |

**Table 2.** Shorthand notations

| Symbol | Meaning |
|---|---|
| $\varepsilon_{ij}$ | $= \left( \nabla_j v_i^s + \nabla_i v_j^s \right)/2 - \nabla_k v_k^s \delta_{ij}/3$, deviator of strain rate |
| $\bar{\rho}$ | $= \rho^f \varphi + \rho^s (1-\varphi)$, total density |
| $\bar{p}$ | $= p^f \varphi + p^s (1-\varphi)$, total pressure |
| $p_e$ | $= \bar{p} - p^f = \left( p^s - p^f \right)(1-\varphi)$, effective pressure |
| $q_i^D$ | $= \varphi \left( v_i^f - v_i^s \right)$, Darcy's flux |
| $\dfrac{d^f}{dt}$ | $= \dfrac{\partial}{\partial t} + v_i^f \nabla_i = \dfrac{d^s}{dt} + \left( v_i^f - v_i^s \right) \nabla_i$, Lagrangian derivative with respect to fluid |
| $\dfrac{d^s}{dt}$ | $= \dfrac{\partial}{\partial t} + v_i^s \nabla_i$, Lagrangian derivatives with respect to solid |